\documentclass[iop]{emulateapj}
\usepackage{natbib,bm,multirow,tabularx}
\usepackage{graphicx, graphics}
\usepackage{threeparttable}
\usepackage{amsmath}
\shortauthors{Chung et al.}
\newcommand{\te}{t_{\rm E}}
\newcommand{\uos}{u_{\rm 0, sat}}
\newcommand{\uoe}{u_{\rm 0, \oplus}}

\newcommand{\pie}{\pi_{\rm E}}
\newcommand{\pien}{\pi_{\rm E,N}}
\newcommand{\piee}{\pi_{\rm E,E}}
\newcommand{\bpie}{\bm{\pi}_{\rm E}}
\newcommand{\bmurel}{\bm{\mu}_{\rm rel}}
\newcommand{\pirel}{\pi_{\rm rel}}
\newcommand{\thetae}{\theta_{\rm E}}
\newcommand{\thetas}{\theta_{\rm \star}}
\newcommand{\uas}{\mu{\rm as}}
\newcommand{\murel}{\mu_{\rm rel}}
\newcommand{\dl}{D_{\rm L}}
\newcommand{\ds}{D_{\rm S}}

\newcommand{\delcs}{\Delta \chi^{2}}

\begin{document}
\title{OGLE-2015-BLG-1482L: the first isolated low-mass microlens in the Galactic bulge }
\author{
S.-J. Chung$^{1,2,17}$, W. Zhu$^{3,17,19}$,  A. Udalski$^{4,18}$, C.-U. Lee$^{1,2,17}$, Y.-H. Ryu$^{1,17}$, Y. K. Jung$^{5,17}$, I.-G. Shin$^{5,17}$, J. C. Yee$^{5,17,19}$, K.-H. Hwang$^{1,17}$, A. Gould$^{1,3,6,17,19}$\\
and\\
M. Albrow$^{7}$, S.-M. Cha$^{1,8}$, C. Han$^{9}$, D.-J. Kim$^{1}$, H.-W. Kim$^1$, S.-L. Kim$^{1,2}$, Y.-H. Kim$^{1,2}$, Y. Lee$^{1,8}$, B.-G. Park$^{1,2}$ , R. W. Pogge$^{3}$\\
(The KMTNet collaboration)\\
R. Poleski$^{3,4}$, P. Mr{\'o}z$^{4}$, P. Pietrukowicz$^{4}$, J. Skowron$^{4}$, M.K. Szyma{\'n}ski$^4$, I. Soszy{\'n}ski$^{4}$, S. Koz{\l}owski$^4$, K. Ulaczyk$^{4,10}$, M. Pawlak$^{4}$\\
(The OGLE collaboration)\\
C. Beichman$^{11}$, G. Bryden$^{12}$, S. Calchi Novati$^{13,14}$, S. Carey$^{15}$, M. Fausnaugh$^{2}$, B.\ S. Gaudi$^{2}$, Calen B. Henderson$^{12,16}$, Y. Shvartzvald$^{12,16}$, B. Wibking$^{2}$\\
(The \textit{Spitzer} team)
 }

\affil{$^1$ Korea Astronomy and Space Science Institute, 776 Daedeokdae-ro, Yuseong-Gu, Daejeon 34055, Korea; sjchung@kasi.re.kr}
\affil{$^2$ Korea University of Science and Technology, 217 Gajeong-ro, Yuseong-gu, Daejeon 34113, Korea}
\affil{$^3$ Department of Astronomy, Ohio State University, 140 W. 18th Ave., Columbus, OH 43210, USA}
\affil{$^4$ Warsaw University Observatory, AI.~Ujazdowskie~4, 00-478~Warszawa, Poland}
\affil{$^5$ Harvard-Smithsonian Center for Astrophysics, 60 Garden St., Cambridge, MA 02138, USA}
\affil{$^6$ Max-Planck-Institute for Astronomy, K{\"o}nigstuhl 17, 69117 Heidelberg, Germany}
\affil{$^7$ Department of Physics and Astronomy, University of Canterbury, Private Bag 4800 Christchurch, New Zealand}
\affil{$^8$ School of Space Research, Kyung Hee University, Giheung-gu, Yongin, Gyeonggi-do, 17104, Korea}
\affil{$^9$ Department of Physics, Chungbuk National University, Cheongju 361-763, Korea}
\affil{$^{10}$ Department of Physics, University of Warwick, Gibbet Hill Road, Coventry, CV4~7AL,~UK}
\affil{$^{11}$ NASA Exoplanet Science Institute, MS 100-22, California Institute of Technology, Pasadena, CA 91125, USA}
\affil{$^{12}$ Jet Propulsion Laboratory, California Institute of Technology, 4800 Oak Grove Drive, Pasadena, CA 91109, USA}
\affil{$^{13}$ IPAC, Mail Code 100-22, Caltech, 1200 E. California Blvd., Pasadena, CA 91125}
\affil{$^{14}$ Dipartimento di Fisica ``E. R. Caianiello'', Universit\`a di Salerno, Via Giovanni Paolo II, 84084 Fisciano (SA), Italy}
\affil{$^{15}$ Spitzer Science Center, MS 220-6,  California Institute of Technology, Pasadena, CA, USA}
\affil{$^{16}$ NASA Postdoctoral Program Fellow}
\affil{$^{17}$ The KMTNet Collaboration}
\affil{$^{18}$ The OGLE Collaboration}
\affil{$^{19}$ The \textit{Spitzer} Team }

% ==================================================================
%\submitted{Submitted to The Astrophysical Journal}

\begin{abstract}
We analyze the single microlensing event OGLE-2015-BLG-1482 simultaneously observed from two ground-based surveys and from \textit{Spitzer}.
The \textit{Spitzer} data exhibit finite-source effects due to the passage of the lens close to or directly over the surface of the source star as seen from \textit{Spitzer}.
Such finite-source effects generally yield measurements of the angular Einstein radius, which when combined with the microlens parallax derived from a comparison between the ground-based and the \textit{Spitzer} light curves, yields the lens mass and lens-source relative parallax.
From this analysis, we find that the lens of OGLE-2015-BLG-1482 is a very low-mass star with the mass $0.10 \pm 0.02 \ M_\odot$ or a brown dwarf with the mass $55\pm 9 \ M_{J}$, which are respectively located at $D_{\rm LS} = 0.80 \pm 0.19\ \textrm{kpc}$ and $ D_{\rm LS} = 0.54 \pm 0.08\ \textrm{kpc}$, where $D_{\rm LS}$ is the distance between the lens and the source, and thus it is the first isolated low-mass microlens that has been decisively located in the Galactic bulge.
The degeneracy between the two solutions is severe ($\delcs = 0.3$).
The fundamental reason for the degeneracy is that the finite-source effect is seen only in a single data point from \textit{Spitzer} and this single data point gives rise to two solutions for $\rho$, the angular size of the source in units of the angular Einstein ring radius.
Because the $\rho$ degeneracy can be resolved only by relatively high cadence observations around the peak, while the \textit{Spitzer} cadence is typically $\sim 1\,{\rm day}^{-1}$, we expect that events for which the finite-source effect is seen only in the \textit{Spitzer} data may frequently exhibit this $\rho$ degeneracy.
For OGLE-2015-BLG-1482, the relative proper motion of the lens and source for the low-mass star is $\murel = 9.0 \pm 1.9\ \textrm{mas\ yr$^{-1}$}$, while for the brown dwarf it is $\murel = 5.5 \pm 0.5\ \textrm{mas\ yr$^{-1}$}$.
Hence, the degeneracy can be resolved within $\sim 10\ \rm yrs$ from direct lens imaging by using next-generation instruments with high spatial resolution.
\end{abstract}
\keywords{brown dwarfs - gravitational lensing: micro - stars: fundamental parameters}

\section{INTRODUCTION}

Microlensing is sensitive to planets orbiting low-mass stars and brown dwarfs (BDs) that are difficult to detect by other methods, such as the radial velocity and transit method.
Although faint low-mass stars such as M dwarfs comprise $\sim 70 \%$ of stars in the solar neighborhood and the Galaxy \citep{skowron15}, it is difficult to detect distant M dwarfs due to their low luminosity.
However, microlensing depends on the mass of the lens, not the luminosity, and thus it is not affected by the distance and luminosity of the lens.
Hence, microlensing is the best method to probe faint M dwarfs in the Galaxy.
A majority of host stars of 52 extrasolar planets detected by microlensing are M dwarfs, and they are distributed within a wide range of distances about $0.4 - 8\ \rm kpc$.

Until now a large number of BDs \citep{han16} have been discovered by various methods including radial velocity \citep{sahlmann11}, transit \citep{deleuil08, johnson11, siverd12, moutou13, diaz13}, and direct imaging \citep{lafreniere07}, and most of them are young \citep{luhman12}.
There exist various scenarios of BD formation based on these plentiful BD samples.
Since microlensing provides different BD samples from other methods, the microlensing BD samples will play an important role to constrain the various BD formation scenarios. 
17 BDs have been detected with microlensing so far.
Only two of them, OGLE-2007-BLG-224L \citep{gould09} and OGLE-2015-BLG-1268L \citep{zhu16}, are isolated BDs, while most of the others, OGLE-2006-BLG-277Lb \citep{park13} , OGLE-2008-BLG-510Lb/MOA-2008-BLG-369L \citep{bozza12}, MOA-2009-BLG-411Lb \citep{bachelet12}, MOA-2010-BLG-073Lb \citep{street13}, MOA-2011-BLG-104Lb/OGLE-2011-BLG-0172 \citep{choi13}, MOA-2011-BLG-149Lb \citep{shin12b}, OGLE-2013-BLG-0102Lb \citep{jung15}, and OGLE-2013-BLG-0578Lb \citep{park15}, are binary companions orbiting M dwarf stars.
This is because binary lens events (i.e., events with anomalies in the light curve) have a larger chance to measure masses of the lens than single lens events, such as isolated BD events.

The key problem in ``detecting" isolated BDs is that in general, we do not know whether they are ``detected" or not, since all that we obtain from observed events is the Einstein timescale $\te$, which is the crossing time of the Einstein radius of the lens.
With the observed $\te$, we can only make a very rough estimate of the lens mass and so cannot distinguish potential BDs from stars.
To measure the masses of isolated BDs in the isolated BDs events, two parameters are required: the angular Einstein radius $\thetae$ and microlens parallax $\pie$.
This is because \citep{gould92, gould00}
\begin{equation}
\label{eqn:mass}
M_{\rm L} = {\thetae\over{\kappa \pie}} 
\end{equation}
and 
\begin{equation}
\pie = {\pirel\over{\thetae}}; \\\\\  \pirel \equiv {\rm AU} \left({{1\over{\dl}} - {1\over{\ds}}}\right),
\end{equation}
where
\begin{displaymath}
\kappa \equiv {4G\over{c^{2}\rm AU}} \approx 8.14{{\rm mas}\over{M_\odot}}.
\end{displaymath}
Here $M_{\rm L}$ is the lens mass, and $D_{\rm L}$ and $D_{\rm S}$ are the distances to the lens and the source from the observer, respectively.
However, it is usually quite difficult to measure the two parameters $\thetae$ and $\pie$.

In general, $\thetae$ is obtained from the measurement of the normalized source radius $\rho = \theta_\star/\thetae$, where $\theta_\star$ is an angular radius of the source.
The $\rho$ measurement is limited to well-covered caustic-crossing events and high-magnification events in which the source passes close to the lens, while $\theta_\star$ is usually well measured through the color and brightness of the source.
Because isolated BD events are almost always quite short, $\pie$ can usually be measured only via so-called terrestrial parallax \citep{gould97, gould09}.
Terrestrial parallax measurements are limited to well-covered high-magnification events.
As a result, it is very hard to measure masses of isolated BDs from the ground \citep{gould13}.  
The best way to measure $\pie$ is a simultaneous observation of an event from Earth and a satellite \citep{refsdal66, gould94b}.
Fortunately, since 2014, the \textit{Spitzer} satellite has been regularly observing microlensing events toward the Galactic bulge in order to measure the microlens parallax.
The \textit{Spitzer} observations suggest a new opportunity to obtain the mass function of BDs from the simultaneous observation from Earth and \textit{Spitzer}, although they are not dedicated to BDs \citep{zhu16}.

The simultaneous observation from the two observatories with sufficiently wide projected separation $D_\perp$ allows to measure the microlens parallax vector $\bpie$ from the difference in the light curves as seen from the two observatories,
\begin{equation}
\bpie = \pie{\bmurel\over{\murel}}, 
\end{equation}
where $\murel$ is the lens-source relative proper motion and
\begin{equation}
\bpie = {{\rm AU}\over{D_\perp}}\left(\Delta \tau, \Delta \beta_{\pm \pm} \right),  
\end{equation}
where
\begin{equation}
\Delta \tau = {{t_{\rm 0,sat} - t_{\rm 0,\oplus}}\over{\te}};\quad \Delta \beta_{\pm \pm} = \pm u_{\rm 0,sat} - \pm u_{\rm 0,\oplus}.
\end{equation}
Here $t_0$ is the time of the closest source approach to the lens (peak time of the event) and $u_0$ is the separation between the lens and the source at time $t_0$ (impact parameter).
The subscripts of $\rm ``sat"$ and $``\oplus"$ indicate the parameters as measured from the satellite and Earth, respectively.
Thus, $\Delta \tau$ and $\Delta \beta$ represent the difference in $t_0$ and $u_0$ as measured from the two observatories.
Parallax measurements made by such comparisons between the light curves are subject to a well-known four-fold degeneracy, which comes from four possible values of $\Delta \beta$ including ($+\uos, \pm \uoe$) and ($-\uos, \pm \uoe$).
However, there is only a two-fold degeneracy in the amplitude of $\bpie$ because $\Delta\beta_{--} = - \Delta\beta_{++}$ and $\Delta\beta_{-+} = - \Delta\beta_{+-}$.
The only exception to the four-fold degeneracy would be if one of the two observatories has $u_0$ consistent with zero, while the other has $u_0$ inconsistent with zero. 
In this case, the four-fold degeneracy reduces to a two-fold degeneracy.
For example, if $u_{0,\rm sat}=0$ (within errors), then $\Delta\beta_{+,+}=\Delta\beta_{-,+}\rightarrow \Delta\beta_{0,+}$ (and similarly for $\Delta\beta_{0,-}$).
Then, since $\Delta\beta_{0,-}=-\Delta\beta_{0,+}$, there is no degeneracy in the mass (See e.g., \citealt{gould12}).  
This case is very important for point-lens mass measurements since the lens always passes very close to or over the source as seen
by one observatory, so $u_0\simeq 0$, whether or not it is strictly consistent with zero.

Here we report the fifth isolated-star measurement derived from microlensing measurements of $\rho$ and $\pie$. 
In contrast to the previous four measurements, this one has a discrete degeneracy in $\rho$ and therefore in mass. 
We trace the origin of this degeneracy to the fact that only a single point is affected by finite-source effects, and we argue that it may occur frequently in future space-based microlensing mass measurements, including BDs.
We show how this degeneracy can be broken by future high-resolution imaging, regardless of whether the lens is dark or luminous.
This paper is organized as follows.
In Section 2, the observation of the event OGLE-2015-BLG-1482 is summarized, and we describe the analysis of the light curve in Section 3.
With the results of Section 3, we derive physical properties of the source and lens in Section 4 and then we discuss the results in Section 5.
Finally, we conclude in Section 6.

\section{OBSERVATIONS}
\subsection{Ground-based observations}

The gravitational microlensing event OGLE-2015-BLG-1482 was discovered by the Optical Gravitational Lensing Experiment (OGLE) (Udalski 2003), and it was also observed by \textit{Spitzer} and Korea Microlensing Telescope Network (KMTNet, \citealt{kim16}).
The microlensed source star of the event is located at ($\alpha$,$\delta$) = ($17^{\rm h}50^{\rm m}31^{\rm s}.33,-30^{\circ}53'19\farcs3$) in equatorial coordinates and ($l$,$b$) = ($358 \fdg 88,-1 \fdg 92$) in Galactic coordinates.

OGLE observations were carried out using a 1.3 m Warsaw telescope with a field of view of 1.4  square degrees at the Las Campanas Observatory in Chile.
The event lies in the OGLE field BLG534 with a cadence of about $0.3\,{\rm hr}^{-1}$ in $I$ band.
The Einstein timescale is quite short, $\te \sim 4\,$days, and the OGLE baseline of this event is slightly variable on long timescales at about the 0.02 mag level.
Thus, we used only 2015 season data for light curve modeling.
 
KMTNet observations were conducted using 1.6 m telescopes with fields of view of 4.0 square degrees at each of three different sites, Cerro Tololo Inter-American Observatory (CTIO) in Chile, the South African Astronomical Observatory (SAAO) in South Africa, and Siding Spring Observatory (SSO) in Australia.
The scientific observations at the CTIO, SAAO, and SSO were initiated on 3 February, 19 February, and 6 June in 2015, respectively.
OGLE-2015-BLG-1482 was observed with $10 - 12$ minute cadence at the three sites and the exposure time was 60 s.
The CTIO, SAAO, and SSO observations were made in $I$-band filter, and for determining the color of the source star, the CTIO observations with a typical good seeing were also made in $V$-band filter.
Thus, the light curve of the event was well covered by the three KMTNet observation data sets.
The KMTNet data were reduced by the Difference Image Analysis (DIA) photometry pipeline \citep{alard98, albrow09}.

\subsection{\textit{Spitzer} observations}

\textit{Spitzer} observations in 2015 were carried out under an 832-hour program whose principal scientific goal was to measure the Galactic distribution of planets \citep{gould14}. 
The event selection and observational cadences were decided strictly by the protocols of \cite{yee15b}, according to which events could be selected either ``subjectively" or ``objectively". 
Events that meet specified objective criteria {\it must} be observed according to a specified cadence. 
In this case all planets discovered, whether before or after \textit{Spitzer }observations are triggered, (as well as all planet sensitivity) can be included in the analysis.
Events that do not meet these criteria can still be chosen ``subjectively".
In this case, planets (and planet sensitivity) can only be included in the Galactic-distribution analysis based on data that become available after the decision. 
Like objective events, events selected subjectively must continue to be observed according to the specified cadence and stopping criteria (although those may be specified as different from the standard, objective values at the time of selection).

Because the current paper is not about planets or planet sensitivities, the above considerations play no direct role. 
However, they play a crucial indirect role. 
Figure 1 shows that despite the event's very short timescale $\te \sim 4\,$days, and despite the fact that it peaked as seen from \textit{Spitzer} slightly before it peaked from Earth, observations began about 1 day prior to the peak. 
This is remarkable because, as discussed in detail by \citet{udalski15} (see their Figure 1), there is a delay between the selection of a target and the start of
the \textit{Spitzer} observations.
Targets can only be uploaded to the spacecraft once per week, and it takes some time to prepare the target uploads.
Therefore, \textit{Spitzer} observations begin a minimum of three days after the final decision is made to observe the event with \textit{Spitzer}, and that decision is generally based on data taken the night before, i.e. about four days prior to the first \textit{Spitzer} observations.
Hence, at the time that the decision was made to observe OGLE-2015-BLG-1482, the source was significantly outside the Einstein ring. 
It is notoriously difficult to predict the future course of such events. 
Therefore, such events cannot meet objective criteria that far from the peak, but selecting them``subjectively" would require a commitment to continue observing them for several more weeks of the campaign, which risks wasting a large number of observations if the event turns out to be very low-magnification with almost zero planet sensitivity (the most likely scenario).
At the same time, if the event timescale is short, it could be over before the next opportunity to start observations with \textit{Spitzer} (~10 days later)

Hence, \citet{yee15b} also specified the possibility of so-called ``secret alerts". 
For these, an observational sequence would be uploaded to \textit{Spitzer} for a given week, but no announcement would be made.  If the event looked promising later (after upload), then it could be chosen subjectively. 
In this case, \textit{Spitzer }data taken after the public alert could be included in the parallax measurement (needed to enter the Galactic-distribution sample) but \textit{Spitzer} data taken before this date could not.
If the event was subsequently regarded as unpromising, it would not be subjectively alerted, in which case the observations could be halted the next week without violating the \citet{yee15b} protocols.

This was exactly the case for OGLE-2015-BLG-1482 (see Figure 1). 
It was $``$secretly" alerted at the upload for observations to begin at HJD$'$ = HJD-2450000 = 7206.73. 
It was only because of this secret alert that an observation was made near peak, which became the basis for the current paper. 
In fact, its subsequent rise was so fast (due to its short timescale) that it was subjectively alerted just prior to the near-peak \textit{Spitzer} observation. 
At the next week's upload, it met the objective criteria. 
Note, however, that if we had waited for the event to become objective before triggering observations, we would not have been able to make the mass measurement reported here, even though the planet sensitivity analysis would have been almost identically the same (provided that parallax could still be measured with the remaining \textit{Spitzer} observations).
This is the first \textit{Spitzer} microlensing event for which a ``secret alert" played a crucial role.  

\textit{Spitzer} observations were made in $3.6\, \mu$m channel on the IRAC camera from HJD$'$ = HJD - 2450000 =  7206.73 to 7221.04.
The data were reduced using specialized software developed specifically for this program \citep{calchinovati15}.
Even though the \textit{Spitzer} data are relatively sparse, there is one point near the peak, which proves to be essential to determine the normalized source radius $\rho$.

\section{LIGHT CURVE ANALYSIS}

Event OGLE-2015-BLG-1482 was densely, and almost continuously covered by ground-based data, but showed no significant anomalies (See Figure 1).
This has two very important implications.  
First, it means that the ground-based light curve can be analyzed as a point lens.
Second, it implies that it is very likely (but not absolutely guaranteed) that the \textit{Spitzer} light curve can likewise be analyzed as a point lens. 
The reason that the latter conclusion is not absolutely secure is that the \textit{Spitzer} and ground-based light curves are separated in the Einstein ring by $\Delta\beta\sim 0.15$.  Thus, even though we can be quite certain that the ground-based source trajectory did not go through (or even near) any caustics of significant size, it is still possible that the source as seen from \textit{Spitzer} did pass through a significant caustic, but that this caustic was just too small to affect the ground-based light curve.

Nevertheless, since the closest \textit{Spitzer} point to peak has impact parameter $u_{spitzer} \sim 0.06$ and it is quite rare for events to show caustic anomalies at such separations when there are no anomalies seen in densely sampled data $u > 0.15$, we proceed under the assumption that the event can be analyzed as a point lens from both Earth and \textit{Spitzer}.
Thus, we conduct the single lens modeling of the observed light curve by minimizing $\chi^2$ over parameter space.
For the $\chi^2$ minimization, we use the Markov Chain Monte Carlo (MCMC) method.
Thanks to the simultaneous observation from the Earth and satellite, we are able to measure the microlens parallax $\pie = (\pi^{2}_{\rm E, N} + \pi^{2}_{\rm E, E})^{1/2}$, which are the north and east components of the parallax vector $\bpie$, respectively.
The \textit{Spitzer} light curve has a point near the peak of the light curve, and thus we can also measure the normalized source radius $\rho$.
Hence, we put three single lensing parameters of $t_0$, $u_0$, and $\te$, the parallax parameters of $\pien$ and $\piee$, and the normalized source radius $\rho$ as free parameters in the modeling.
In addition, there are two flux parameters for each of the 5 observatories (\textit{Spitzer}, OGLE, KMT CTIO, KMT SAAO, KMT SSO).
One represents the source flux $f_{s,i}$ as seen from the $i$th observatory, while the other, $f_{b,i}$  is the blended flux within the aperture that does not participate in the event. 
That is, the five observed fluxes $F_i(t_j)$ at epochs $t_j$ are simultaneously modeled by
\begin{equation}
F_i(t_j) = f_{s,i}A_i(t_j;t_0,u_0,\te,\rho,\bpie) + f_{b,i},
\label{eqn:ftot}
\end{equation}
where $A_i(t)$ is the magnification as a function of time at the $i$th observatory.
In principle, these magnifications may differ because the observatories are at different locations. 
However, in this event the separations of the observatories on Earth are so small compared to the projected size of the Einstein ring that we ignore them and consider
all Earth-based observations as being made from Earth’s center.
That is, we ignore so-called ``terrestrial parallax".
At the same time, the distance between the Earth and \textit{Spitzer} remains highly significant, so $A_{Spitzer}(t)$ is different from $A_{\rm Earth}(t)$.
As is customary (e.g., \citealt{dong07, udalski15, yee15a}), we determine the parameters in the $``$geocentric" frame at the peak of the event as observed from Earth  \citep{gould04}, and likewise adopt the sign conventions shown in Figure 4 of \citet{gould04}.

In addition, we conduct the modeling for the point-source/point-lens, because only a single point of \textit{Spitzer} contributes to the finite-source effect.
We find that the $\delcs$ between the best-fit models of the point- and finite-sources is $\delcs = 31.47$.
Hence, OGLE-2015-BLG-1482 strongly favors the finite-source model.

\subsection{Limb Darkening}

As we will show, the lens either transits or passes very close to the source as seen by \textit{Spitzer}, which induces finite-source effects near the peak of the \textit{Spitzer} light curve. 
To account for this, we adopt a limb-darkened brightness profile for the source star of the form
\begin{equation}
S_{\lambda}(\theta) = \bar{S}_{\lambda}\left[1 - \Gamma\left(1 - {3\over{2}}\cos\theta\right)\right],
\end{equation}
where $\bar{S}_{\lambda} \equiv F_{S, \lambda}/(\pi \theta^{2}_{\star})$ is the mean surface brightness of the source, $F_{S,\lambda}$ is the total flux at wavelength $\lambda$, $\Gamma$ is the limb darkening coefficient, and $\theta$ is the angle between the normal to the surface of the source star and the line of sight \citep{an02a}.
Based on the estimated color and magnitude of the source, which is discussed in Section 4, assuming an effective temperature $T_{\rm eff} = 4500\ \rm K$, solar metallicity, surface gravity $\log\ g = 0.0$, and microturbulent velocity $v_{t} = 2$ km/s, we adopt $\Gamma_{3.6\,\mu {\rm m}} = 0.178$ from \citet{claret11}.

\subsection{$(2\times 2)=4$ highly degenerate solutions}

As discussed in Section 1, space-based parallax measurements for point lenses generically give rise to four solutions, which can be highly degenerate. 
However, in cases for which one of two observations has $u_0 \simeq 0$, while the other has $u_0 \neq 0$, the four solutions reduce to two solutions.
Since for event OGLE-2015-BLG-1482, $Spitzer$ has $\uos \simeq 0$, we expect the event to have two degenerate solutions, $\uoe > 0$ and $\uoe < 0$.
However, what we see in Table 1 is not two degenerate solutions but four.
For each of the two expected degenerate solutions [$(+,0), (-,0)$], there are two solutions with different values of $\rho$ ($\rho \simeq 0.06$ and $\rho \simeq 0.09$).
Figure 1 shows the best-fit light curve of the event OGLE-2015-BLG-1482 with the OGLE, KMT, and \textit{Spitzer} data sets.
The best-fit solution is $(+,0)$ solution for $\rho \simeq 0.06$, which means  $\uoe > 0$ and $\uos \simeq 0$.
The biggest $\delcs$ between the four solutions is $\delcs \simeq 0.5$.

We should expect the two-fold parallax to be very severe in this case.
This two-fold degeneracy would be exact in the approximations that 1) Earth and \textit{Spitzer} are in rectilinear motion and 2) they have zero relative projected velocity \citep{gould95}. 
For events that are very short compared to a year (like this one), the approximation of rectilinear motion is excellent. 
And while Earth and \textit{Spitzer} had relative projected motion of order $v_\oplus \sim 30\,{\rm km\,s^{-1}}$, this must be compared to the lens-source projected velocity $\tilde v$,
\begin{equation}
\tilde v \equiv {{\rm AU}\over \pie\te}\simeq 3050\,{\rm km\,s^{-1}}.
\label{eqn:tildev}
\end{equation}
Hence, these two solutions are almost perfectly degenerate.
On the other hand, the $\rho$ degeneracy was completely unexpected.
It is also very severe.
The origins of the $\rho$ degeneracy are discussed in Section 5.
To illustrate the $\rho$ degeneracy, the light curve of the best-fit model $(+,0)$ for $\rho \simeq 0.09$ is also presented in Figure 1.
In Table 1, we present the parameters of all the four solutions.

\section{Physical properties}
\subsection{Source properties}
The color and magnitude of the source are estimated from the observed $(V - I)$ source color and best-fit modeling of the light curve, but they are affected by extinction and reddening due to the interstellar dust along the line of sight.
The dereddened color and magnitude of the source can be determined by comparing to the color and magnitude of the red clump giant (RC) under the assumption that the source and RC experience the same amount of reddening and extinction \citep{yoo04}.
Figure 2 shows the instrumental KMT CTIO color-magnitude diagram (CMD) of stars in the observed field.
The color and magnitude of the RC are obtained from the position of the RC on the CMD, which correspond to $[(V - I), I]_{\rm RC} = [1.67, 17.15]$.
We adopt the intrinsic color and magnitude of the RC with $(V - I)_{\rm RC,0}$ = 1.06 \citep{bensby11} and $I_{\rm RC,0}$ = 14.50 \citep{nataf13}.
The instrumental source color obtained from a regression is $(V - I)_{\rm s} = 1.74$ and the magnitude of the source obtained from the best-fit model is $I_{\rm s} = 17.37$.
The measured offset between the source and the RC is $[\Delta(V - I), \Delta I] = [0.07, 0.22]$.
Here we note that there exists an offset between the instrumental magnitudes of OGLE and KMTNet as $I_{\rm kmt} - I_{\rm ogle} = 0.045$ mag.
Thus, we should consider the offset when we estimate the dereddened magnitude of the source.
As a result, we find the dereddened color and magnitude of the source $[(V - I), I]_{\rm s,0} = [1.13, 14.76]$.
The dereddened $(V - K)$ source color by using the color-color relation of \citet{bessell88} is $(V - K)_{\rm s,0} = 2.61$.
Then adopting $(V - K)_{\rm s,0}$ to the  the color-surface brightness relation of \citet{kervella04}, we determine the source angular radius $\thetas = 5.79 \pm 0.39\ \uas$.
The estimated color and magnitude of the source suggest that the source is a K type giant.
The error in $\thetas$ includes the uncertainty in the source flux, the uncertainty in the conversion from the observed $(V - I)$ color to the surface brightness, and the uncertainty of centroiding the RC.
The uncertainty in the source flux is about $1\%$ and the uncertainty of the microlensing color is $0.02$ mag, which contributes $1.6\%$ error in $\thetas$ measurement.
The scatter of the source angular radius relation in $(V - K)_{\rm s,0}$ is $5\%$ \citep{kervella08}, and centroiding the RC contributes $4\%$ to the radius uncertainty \citep{shin16}.

As mentioned above, since the degeneracy between two different $\rho$ solutions is very severe as $\delcs \lesssim 0.3$, we should consider both $\rho$ solutions. 
The two $\rho$ values yield two different Einstein radii, 
\begin{equation}
\thetae = \thetas/\rho = \left\lbrace 
\begin{array}{ll}
0.104 \pm 0.022\ \textrm{mas} & \textrm{for $\rho \simeq 0.06$} \\\\
0.063 \pm 0.006\ \textrm{mas} & \textrm{for $\rho \simeq 0.09$}.
\end{array} \right.
\end{equation}
Because of the two different Einstein radii, all the physical parameters related to the lens take on two discrete values.
The relative proper motions of the lens and source are,
\begin{equation}
\murel = \thetae/\te = \left\lbrace 
\begin{array}{ll}
8.96 \pm 1.88\ \textrm{mas\ yr$^{-1}$} & \textrm{for $\rho \simeq 0.06$} \\\\
5.48 \pm 0.48\ \textrm{mas\ yr$^{-1}$} & \textrm{for $\rho \simeq 0.09$}.
\end{array} \right.
\end{equation}

\subsection{Lens properties}

The mass and distance of the lens can be obtained from the measured Einstein radius $\thetae$ and microlens parallax $\pie$.
As discussed in the introduction, the four-fold degeneracy in $\bpie$ usually leads to a two-fold degeneracy in its amplitude $\pie$.
However, in the case of events that are much higher magnification (much lower $u_0$) as seen from one observatory than the other, the two-fold degeneracy collapses as well. 
This is because, under these conditions, $|\Delta\beta_{\pm\pm}|\simeq |\Delta\beta_{\pm\mp}|$. 
The present case is consistent with the lens passing exactly over the center of the source as seen by \textit{Spitzer} (to our ability to measure it). 
Then, according to Equation \eqref{eqn:mass}, we measure the lens mass, 
\begin{displaymath}
M = {\thetae\over{\kappa \pie}} = \left\lbrace
\begin{array}{ll}
0.096 \pm 0.023 \ M_\odot & \textrm{for $\rho \simeq 0.06$} \\\\
0.055 \pm 0.009 \ M_\odot & \textrm{for $\rho \simeq 0.09$}.
\end{array} \right.
\end{displaymath}
The lens-source relative parallax for the two cases is
\begin{equation}
\pi_{\rm rel} = \thetae \pie = \left\lbrace
\begin{array}{ll}
0.014 \pm 0.003\ {\rm mas} & \textrm{for $\rho \simeq 0.06$} \\\\
0.009 \pm 0.001\ {\rm mas} & \textrm{for $\rho \simeq 0.09$}.
\end{array} \right.
\end{equation}
These values of $\pi_{\rm rel}$ are very small compared to the source parallax $\pi_s \sim 0.12\,$mas. 
This implies that the distance between the lens and the source is determined much more precisely than the distance to the lens or the source separately. 
That is,
\begin{equation}
D_{\rm LS} \equiv D_{\rm S} - D_{\rm L} =  {{\pirel \over {\rm AU}} D_{\rm S} D_{\rm L}}
\label{eqn:dls}
\end{equation}
\begin{displaymath}
 \simeq \left\lbrace
\begin{array}{ll}
0.80 \pm 0.19 \ {\rm kpc} & \textrm{for $\rho \simeq 0.06$} \\\\
0.54 \pm 0.08 \ {\rm kpc} & \textrm{for $\rho \simeq 0.09$}.
\end{array} \right.
\end{displaymath}

Since the source is almost certainly a bulge clump star (from its position on the CMD), and the lens is  $\lesssim 1$ kpc from the source, it is likewise almost certainly in the bulge. 
Thus, this is the first isolated low-mass object that has been determined to lie in the Galactic bulge.

\section{Discussion}
\subsection{Future Resolution of the $\rho$ Degeneracy Using Adaptive Optics}

Event OGLE-2015-BLG-1482 has a very severe two-fold degeneracy in $\rho$, in which the $\delcs$ between the two solutions ($\rho \simeq 0.06$ and $\rho \simeq 0.09$) is $\delcs \sim 0.3$.
For the solutions with $\uoe > 0$ and $\uoe < 0$, the microlens parallax vectors $\bpie$ are different from one another, but they have almost the same amplitude $\pie$.
Therefore, the two solutions yield almost the same physical parameters of the lens.
However, each of the two solutions also has two degenerate $\rho$ solutions: $\rho \simeq 0.06$ and $\rho \simeq 0.09$.
Each $\rho$ solution yields different physical parameters of the lens, in particular the lens mass.
For $\rho \simeq 0.06$, the lens is a very low-mass star, while for $\rho \simeq 0.09$ it is a brown dwarf.
The degeneracy of the lens mass due to the two $\rho$ can be resolved from direct lens imaging by using instruments with high spatial resolution (Han \& Chang 2003; Henderson et al. 2014), such as the VisAO camera of the 6.5m Magellan telescope with the resolution $\sim 0\farcs04$ in the $J$ band \citep{close13}\footnote{Close et al. (2014) have obtained a diffraction limited FWHM in ground-based 6m $R$ band images, which gives hope for optical AO. However, it is premature to claim that this technique can be applied to faint stars in the Galactic bulge} and the GMTIFS of the 24.5 m Giant Magellan Telescope (GMT) with resolution $\sim 0\farcs 01$ in the NIR \citep{mcgregor12}.
In general, direct imaging requires 1) that the lens be luminous, and 2) that it be sufficiently far from the source to be separately resolved. 
In the present case, (1) clearly fails for the BD solution.
Hence, the way that high-resolution imaging would resolve the degeneracy is to look for the luminous (but faint) M dwarf predicted by the other solution at its predicted orientation (either almost due north or due south of the source -- since $|\pi_{\rm E,N}|\gg |\pi_{\rm E,E}|$) and with its predicted separation $(t_{\rm AO} - 2015)\times (9\,\rm mas\,yr^{-1})$.
If the M dwarf fails to appear at one of these two expected positions, the BD solution is correct. 
Since the source is a clump giant, and hence roughly $10^4$ times brighter than the M dwarf, it is likely that the two cannot be separately resolved until they are separated by at least 2.5 FWHM. 
This requires to wait until $2015 + 2.5\times(40/9)=2026$ for Magellan or $2015 + 2.5\times (10/9)=2018$ for GMT.

\subsection{Origin of the $\rho$ Degeneracy}
The degeneracy in $\rho$ was completely unexpected. 
Indeed we discovered it accidentally because $\rho$ had one value in one of two degenerate parallax solutions and the other value in another one. 
Originally, this led us to think that it was somehow connected to the parallax degeneracy. 
However, by seeding both solutions with both values of $\rho$ we discovered that it was completely independent of the parallax degeneracy.

In retrospect, the reason for this degeneracy is ``obvious''.
There is only a single point that is strongly impacted by the finite size of the source. 
The value of $u$ at this time is well predicted by the rest of the light curve, in particular because \textit{Spitzer} data begin before peak (see Section~2),
\begin{equation}
u = \sqrt{\tau^2 + u_0^2} \quad{\mathrm where} \quad \tau = \frac{(t-t_{\rm 0, sat})}{t_{\rm E}}.
\end{equation}
Hence, the magnification (for point-lens/point-source geometry in a high magnification event) is also known $A_{\rm ps}\simeq 1/u$. 
Moreover, both $f_s$ and $f_b$ for \textit{Spitzer} are also well measured, so that the measured flux at the near-peak point $F$ directly yields an empirical magnification $A_{\rm obs} = (F-f_b)/f_s$ (i.e. the magnification in the presence of finite-source effects). 
Following \citet{gould94a}, the ratio of $A_{\rm obs}$ and $A_{\rm ps}$ can therefore be derived directly from the light curve
\begin{equation}
\label{eqn:bofz}
B(z) \equiv {A_{\rm obs}\over A_{\rm ps}}\simeq A_{\rm obs}u .
\qquad (z\equiv u/\rho)
\end{equation}

As shown by Figure 1 of \citet{gould94a}, $B(z)$ reaches a peak at $z\simeq 0.91$, with $B=1.34$.\footnote{While Figure 1 from \citet{gould94a} shows the correct qualitative behavior, it has a quantitative error in that the peak is at ~1.25, rather than 1.34 (the correct value)} 
Therefore, if one inverts a measurement of $B(z)$ to infer a value of $z$, there are respectively one, two and zero solutions for $B_{\rm obs}<1$, $1<B_{\rm obs}<1.34$, and $B_{\rm obs}>1.34$.

Since this event is a high-magnification event only for \textit{Spitzer}, i.e., the finite-source effect is only seen by \textit{Spitzer}, only the trajectory of \textit{Spitzer} is considered.
Figure 3 (adapted from \citealt{gould94a}) shows the finite-source effect function $B(z)$ as a function of $z$.
For this event, $B(z) = A_{\rm obs}u = 19.14\times0.06 = 1.15$ at the nearest point to the peak, which is indicated by the horizontal dotted line in the figure.
As shown in Figure 3, the function $B(z) = 1.15$ is satisfied at two different values of $z = 0.64$ and $z = 1.12$, which implies (as outlined above) that there are two $\rho$ values.
The two $z$ values yield two normalized source radii of $\rho = 0.094$ (for $z=0.64$) and $\rho = 0.054$ (for $z=1.12$). 
These two derived $\rho$ values are almost the same as those obtained numerically from the best-fit solutions.
Because high-magnification events can be alerted in real time, the high-magnification events observed from Earth are often well covered around the peak by intensive follow-up observations, and thus $\rho$ is almost always well measured if there are significant finite-source effects (i.e., $B\not=1$ for some points).
This means that the $\rho$ degeneracy will often be resolved in high-magnification events observed from the ground.
On the other hand, since the observation cadence of \textit{Spitzer} is much lower than those of ground-based observations, the $\rho$ degeneracy can occur frequently in high-magnification events observed by \textit{Spitzer}.
Note that, in contrast to Figure 1 of \citet{gould94a}, our Figure 3 shows $B(z)$ with and without the effects of limb darkening. 
The effect is hardly distinguishable by eye, in particular because limb darkening at $3.6\,\mu$m is very weak.
Nevertheless this effect should be included.

If finite-source effects are reliably detected from a single measurement near peak, how often will $\rho$ be ambiguous, and if it is ambiguous, how often will the value fall in the upper versus lower allowed ranges?
We might judge there to be a``reliable detection" of finite-source effects from a single point if $|B-1| > X$, where $X$ might be taken as 5\%.
For high-magnification events including the limb darkening effect, we can Taylor expand $B$ for $z>1$ (see Appendix)
\begin{equation}
B(z) = 1 + {1\over{8}}\left(1 - {\Gamma\over5}\right){1\over{z^2}} + {3\over 64}\left(1 - {11\over 35}\Gamma\right){1\over{z^4}} + \ldots,
\end{equation}
where $\Gamma$ is the limb darkening coefficient, as mentioned in Section 3.
Truncating at the second term, we have $B(z) \simeq 1 + (1 - \Gamma/5)/(8z^{2})$.
For \textit{Spitzer} $\Gamma/5 \ll 1$, so we can ignore it. 
Then $B(z) = 1 + 1/(8z^2)$. 
Thus, $B - 1 = X$, i.e., $B = 1 + X$, implies $z = (1/(8X))^{1/2} \rightarrow 1.6$ (for $X = 5\%$).
To next order, $z=(4/3(\sqrt{1+12X}-1))^{-1/2}=1.685$ which is very close to the numerical value, 1.7.
Hence, we have three ranges of recognizable finite-source effects.
The ranges are presented in Table 3.
Table 3 shows that $0.51/(0.51 + 0.34 + 0.79) = 31\%$ of the finite-source effects will be unambiguous. 
And of the times they are ambiguous $0.34/(0.34+0.79) = 30\%$ will have the higher value of $\rho$.

Figure 4 shows the $\chi^2$ distribution of $\uos$ versus $\rho$ from the MCMC chains of the four degenerate solutions in Table 1.
The figure shows that the distribution is centered on $\uos = 0.0$ and thus the four solutions are consistent with $\uos = 0.0$, although there is scatter.
Therefore, it is correct to label $\uos$ as ``0".
The figure also shows that the nearest point to the peak of \textit{Spitzer} light curve favors $\uos =0$, but can accommodate other values of $\uos$, up to about 0.03 at $< 2\sigma$.
In this case,  the bigger $\uos$ makes $B(z)$ bigger because $B(z) = uA_{obs}$, and so allowing values of $z$ between the two best-fit values. 
At the nearest point to the peak, $\tau = |(t - t_{0,sat})|/\te = |(7207.50 - 7207.76)|/4.26 = 0.06$.
Then, 
\begin{eqnarray}
\frac{B(\uos=0.03)}{B(\uos=0)} & = & \frac{u(\uos=0.03)}{u(\uos=0)} \nonumber \\
& = & {\sqrt{{0.06}^2 + 0.03^2\over{0.06^2}}} = 1.12
\end{eqnarray}
Since $B(\uos=0) = 1.15$ from Figure 3, $B(\uos=0.03) = 1.15\times1.12 = 1.29$, and it is the maximum value allowed $B$, and thus the maximum allowed $\uos$.
The allowed maximum $B(z) = 1.29$ yields $z=0.79$ and $z=0.98$ and hence two $\rho$ values, $\rho = 0.085$ and $\rho = 0.068$.
Thus, $\rho \simeq 0.09$ solutions have the lower limit of $\rho =0.085$, while $\rho \simeq 0.06$ solutions have the upper limit of $\rho = 0.068$.

\subsection{$\rho$ degeneracy of OGLE-2015-BLG-0763}

OGLE-2015-BLG-0763 is the only other event with a single lens mass measurement based on finite-source effects observed by \textit{Spitzer} \citep{zhu16}. 
As with OGLE-2016-BLG-1482, the \textit{Spitzer} light curve shows only one point that is strongly affected by finite-source effects (i.e., $B\not=1$).
\citet{zhu16} report $\rho=0.0218$, $\te=33\,$days and their solution implies $t_{0,\rm sat}\simeq 7188.60$ and $\uos=0.012$.
Therefore, the three points closest to peak \citep{calchinovati15} at $t-7188.60 = (-0.75,0.36,0.72)\,$days, have respectively, $u=(0.026, 0.016, 0.025)$.
Since the measurement was derived primarily from the highest point, one may infer $z\equiv u/\rho=0.73$.
Inspection of Figure 3 shows that this implies $B(z)=1.25$, which (since $B>1$) implies that there is another solution at $z=1.01$ and therefore with $\rho=0.016$. 
We can then derive for the two solutions $z=(1.19,0.73,1.15)$ (adopted) and $z=(1.62,1.01,1.57)$ (other).
These yield values of $B$ (from Figure~3) of $B(z)=(1.13,1.25,1.14)$ (adopted) and $B(z)=(1.06,1.25,1.06)$ (other).
That is, for OGLE-2015-BLG-0763, the two nearest points to the peak will both be about 0.08 mag brighter in the adopted solution than in the other solution.
Since the \textit{Spitzer} photometric errors are small compared to these inferred differences (Figure~2 of \citealt{zhu16}), we expect that, in the case of OGLE-2015-BLG-0763 (and in contrast to OGLE-2015-BLG-1482), the near-peak points resolve the degeneracy between the two solutions.

Armed with the above understanding, which was derived without any detailed modeling, we reanalyze OGLE-2015-BLG-0763 and find only an upper limit of 0.01 for the second $\rho$.
However, as discussed in \citet{zhu16}, solutions of the second $\rho$ result in inconsistency with observations, and thus they are not physically correct. 
As a result, there is no $\rho$ degeneracy for the event OGLE-2015-BLG-0763.
As mentioned before, this is because of the near-peak points.
This implies that although for events in which the finite-source effect is seen only in the \textit{Spitzer} the $\rho$ degeneracy can occur frequently due to low observation cadence of the \textit{Spitzer}, it can be resolved by a few data points near the peak.

\subsection{Error analysis in $\rho$ measurement}

The error in the $\rho$ measurement of the event OGLE-2015-BLG-1482 is $19.8\%$ for $\rho \simeq 0.06$ and $6.6\%$ for $\rho \simeq 0.09$.
These errors are quite big relative to measurements in high-magnification events from the ground.
We therefore study the source of these errors in $\rho$ both to determine why they are so different and to make sure that we are properly incorporating all sources of error in this measurement.

As outlined above, the train of information is basically captured by $\rho= u/z(B)$ where $z(B)$ is the inverse of $B(z)$ and both $u$ and $B$ can be regarded approximately as ``measured"
quantities. 
It is instructive to further expand this expression
\begin{equation}
\rho = {u\over z (A_{\rm obs}u)}.
\label{eqn:rho}
\end{equation}
In this form, it is clear that the contribution from an error in determining $u$ tends to be suppressed if $z'\equiv dz/dB>0$ (i.e., $z<0.91$, so $\rho \simeq 0.09$ in our case), and it tends to be enhanced if $z'<0$. 
Hence, this feature of Equation \eqref{eqn:rho} goes in the direction of explaining the larger error in the $\rho \simeq 0.06$ case.
Second, if we for the moment ignore the error in  $u$, then Equation \eqref{eqn:rho} implies $\sigma(\ln\rho) =|z'/z|\sigma(A_{\rm obs})$.
For the two cases, $\rho=(0.09,0.06)$, we have $z=(0.64,1.12)$, $z'=(0.85,-2.18)$ and so $|z'/z|=(1.33,1.95)$. 
Hence, this aspect also favors larger errors for the smaller $\rho$ (larger $z$) solution.
This is intuitively clear from Figure 3: the shallow slope of $B(z)$ toward large $z$ makes it difficult to estimate $z$ from a measurement of $B$.
Hence, the fact that the fractional error in $\rho$ is much larger for the large $z$ (small $\rho$) solution is well understood.

Ignoring blending, we can write $A_{\rm obs} = F/f_s$. 
The error in $F$ (i.e., the flux measurement at the high point) is uncorrelated with any other error.
Since in our case, $u_{0,spitzer}\simeq 0$, we can write $u=(t-t_0)/t_{\rm E}$, and so
\begin{equation}
B = A_{\rm obs}u = {|t-t_0|F\over f_s t_{\rm E}}.
\label{eqn:aobs}
\end{equation}
Since $t_0$ is known extremely well, and $t$ is known essentially perfectly, there would appear to be essentially no error in $|t-t_0|$.
The denominator is a near-invariant in high-magnification events \citep{yee12}. 
That is, the errors in this product are generally much smaller than the errors in either one separately. 
This means that the error in $B$ (and so $z(B)$) is dominated by the flux measurement error of the single point that is affected by finite-source effects.

Nevertheless, it is important to recognize that \citet{yee12} derived their conclusion regarding the invariance of $f_s t_{\rm E}$ under conditions that the error in $f_s$ is completely dominated by the model, and not by the flux measurement errors.
Indeed, as a rather technical, but very relevant point, it is customary practice to ignore the role of flux measurement errors in the determination of $f_s$. 
That is, $f_s$ and $f_b$ are normally {\it not} included as chain variables when modeling microlensing events. 
Instead, the magnification is determined at each point along the light curve from microlens parameters that are in the chain, and then the two flux parameters (from each observatory) are determined from a linear fit. 
This is a perfectly valid approach for the overwhelming majority of microlensing events because the errors arising from this fit (which are returned but usually not reported from the linear fit routine) are normally tiny compared to the error in $f_s$ due to the model. 
Moreover, there are usually many observatories contributing to the light curve, and if all the flux parameters were incorporated in the chain, it would increase the convergence time exponentially.

However, in the present case $t_{\rm E}$ is essentially determined from ground-based data, which are both numerous and very high precision, while $f_s$ is determined from just 16 \textit{Spitzer} points (i.e., all the points save the one near peak).
 If the usual (linear fit) procedure were applied, it would seriously underestimate the error in $f_s$ and so overestimate its degree anticorrelation with $t_{\rm E}$.
We therefore include $(f_s,f_b)_{spitzer}$ as chain parameters and remodel this event.
The result of the remodeling is presented in Table 2.
By comparing to runs in which we treat these flux parameters in the usual way, we find that including these parameters in the chain contributes about 41\% to the $\rho$ error compared to all other sources of $\rho$ error combined.
That is, in the end, this does not dramatically increase the final error, since $(1^2 + 0.41^2)^{1/2}=1.08$. 
Nevertheless, it is important to treat $(f_s,f_b)_{spitzer}$ in a formally proper way since this contribution could easily be the dominant
one in other cases.

\subsection{Impact of the $\rho$ Degeneracy}

The $\rho$ degeneracy was not realized until now for several reasons. 
First of all, although single lens finite-source events have been routinely detected from ground-based observations, they are not scientifically very interesting without the measurement of $\pie$.
However, $\pie$ measurements of single lens events based on ground-based data alone are intrinsically rare and technically difficult \citep{gould13}. 
Second, prior to the establishment of second-generation microlensing surveys, observations of high-magnification microlensing events were usually conducted under the survey+followup mode, which was first suggested by \citet{gould92b}.
High-magnification events with their nearly 100\% sensitivity to planets \citep{griest98} were therefore often followed up with intensive ($\sim$1 min cadence) observations, which could easily resolve this $\rho$ degeneracy, if it exists.

The $\rho$ degeneracy is nevertheless important for the science of second-generation ground-based and future space-based microlensing surveys.
The majority of events found by these surveys will not be followed up at all, and thus the $\rho$ degeneracy can appear because the typical source radius crossing time, $t_\star$, is comparable to the observing cadences that these surveys adopt.
Here
\begin{equation}
\label{eqn:tstar}
t_\star \equiv \frac{\thetas}{\murel} = 45\ {\rm min} \left(\frac{\thetas}{0.6~\mu as}\right) \left(\frac{\murel}{7~\ {\rm mas\ yr^{-1}}}\right)^{-1}\ ,
\end{equation}
where $0.6~\mu$as is the angular source size of a Sun-like star in the Bulge, and 7 $\rm mas\ yr^{-1}$ is the typical value for lens-source relative proper motion of disk lenses.
For second-generation microlensing surveys like OGLE-IV and KMTNet, although a few fields are observed once every $<20$ min, the majority of fields are observed at $>1$ hr cadences. 
Therefore, the single lens finite-source events in these relatively low-cadence fields are likely to have one single data point probing the finite-source effect, and thus the $\rho$ degeneracy appears. 

Fortunately, however, the result of event OGLE-2015-BLG-0763 showed that a few additional data points (before/after crossing the source) around the peak play a crucial role in resolving the $\rho$ degeneracy.
When we observe typical microlensing events with a cadence of 1 hr, we can obtain 2 more data points right before and after crossing the source, except one source-crossing data point.
In this case, the $\rho$ degeneracy will be resolved as in the event OGLE-2015-BLG-0763.
This implies that 1 hr is the upper limit of the observing cadence to resolve the $\rho$ degeneracy in typical single lens events to be observed from the second-generation ground-based surveys, whereas for events with high $\murel$, such as events caused by a fast moving lens object or a high-velocity source star, it is not enough to resolve the $\rho$ degeneracy.

Since about half of KMTNet fields have $\leqslant 1$ hr cadences and these fields have higher probability of detecting events than the other fields with cadences of $\geqslant 2.5$ hr, the $\rho$ degeneracy will be resolved in the majority of single high-magnification events to be observed by KMTNet.
Although $\pie$ is still intrinsically hard to measure even with second-generation surveys, the fraction of events with finite-source effects can be used as an indicator of the properties of the lens population, which is especially important for validating the short-timescale events such as the population of free-floating planets (FFPs) \citep{sumi11}.

\textit{The Wide-Field InfraRed Survey Telescope} (\textit{WFIRST}) is likely to have six microlensing campaigns, each with 72 days observing a $\sim$3 deg$^2$ microlensing field at 15 min cadence \citep{spergel15}. 
\textit{WFIRST} microlensing is expected to detect thousands of bound planets and hundreds of FFPs.
At first sight, the 15 min cadence that \textit{WFIRST} microlensing is currently adopting suggests that it will not be affected by the $\rho$ degeneracy.
However,  since \textit{WFIRST} will go much fainter than ground-based surveys, most of the sources for \textit{WFIRST} events will be M dwarfs, which are a half or even a quarter the size of Sun.
Then the typical $t_\star$ for \textit{WFIRST} events is $\sim$15 min, which is the same as the adopted cadence.
Hence, as mentioned above, although the $\rho$ degeneracy can be usually resolved by obtaining more than 2 data points around the peak, it will be severe for a significant fraction of events with high $\murel$.
What makes this $\rho$ degeneracy more important for \textit{WFIRST} is that $\pie$ can be measured relatively easily once ground-based observations are taken simultaneously \citep{yee13, zhu16b}.
Therefore, the degeneracy in $\rho$ will directly lead to a degeneracy in the mass determination of isolated objects, including FFPs, BDs, and stellar-mass black holes.

\subsection{Potential for the second body}

As discussed in Section 3, it is possible that the deviation of the highly magnified \textit{Spitzer} point might be caused by a caustic structure rather than being entirely due to finite-source effects.
It is easy to show qualitatively that this could affect the exact nature of the system, but is unlikely to significantly change the conclusion that the lens is a low-mass object in the bulge.
First, if there were a caustic perturbation, there would be a second body in the lens system.
However, we do not see any evidence for the second body in the ground-based light curve, and therefore the dominant lensing effect still comes from a single star.
Second, consider the effect on the inferred physical properties of the lens (e.g., mass and distance).
If there were a caustic structure, then it is likely to be small since it does not affected the ground-based data.
In that case $\rho$ would be smaller and therefore $\thetae$ would be larger.
At the same time, $\te$ is clearly determined from the dense, ground-based observations, so if $\thetae$ is larger, $\murel$ must also be larger.
However, $\murel$ is already 9 $\rm mas\ yr^{-1}$ for the smaller $\rho$ solution.
Larger values of $\murel$ are increasingly improbable and will eventually become unphysical.
Hence, OGLE-2015-BLG-1482 is likely an event caused by the single lens star.

However, since a binary lens system could simultaneously reproduce the single lens-like light curve from ground-based observations and the poorly sampled light curve from \textit{Spitzer}, we conduct binary lens modeling.
As a result, we find that the best-fit binary lens solution is the BD binary lens system composed of a primary star $M_{\rm L,1}=0.06 \pm 0.01\ M_\odot$ and a secondary star $M_{\rm L,2}=0.05 \pm 0.01\ M_\odot$ with their projected separation 19 AU, which correspond to lensing parameters of the mass ratio between the binary components $q=0.78$ and the projected separation in units of $\thetae$ of the lens system $s=24$.
The estimated distance to the BD binary is 7.5 kpc, and thus it is also located in the Galactic bulge.
Although $\chi^2$ of the binary lens model is smaller than that of the single lens model by 35, it is a very wide binary system with large $\rho = 0.066$, and thus it is extremely closely related to (in fact, a variety of) the single lens solution with $\rho \simeq 0.06$.

It is important to understand the reason for this close relation.
The key point is that the high point of \textit{Spitzer} is explained by finite-source effects on the tiny caustic of the very wide BD binary that ``replaces'' the point caustic of the point lens in the single lens solution.
But the $\chi^2$ improvement for the binary solution comes entirely from ground-based data, while the $\chi^2$ of the \textit{Spitzer} becomes slightly worse than that of the single lens model (see Figure 1).
Thus, the \textit{Spitzer} high point is not caused by the binary, as we originally sought to test.
The $\chi^2$ improvement could in principle be due to  a distant companion.
However, low-level systematics can also easily produce $\delcs = 35$ improvements in microlensing light curves, which could then mistakenly be attributed to planets, binaries, etc.
For this reason, \citet{gaudi02} and \citet{albrow01} already set a threshold at $\delcs \geqslant 60$ for the detection of a planet based on experience with several dozen carefully analyzed events.
Thus, all we can say about OGLE-2015-BLG-1482 is that the lens is consistent with being isolated but that we cannot rule out that it has a distant companion.
And that the ``evidence" for such a companion is consistent with the systematic effects often seen in microlensing events.

In order to find out whether there is a binary solution for which the high point of \textit{Spitzer} is actually explained by the caustic of a binary, we also conduct binary lens modeling in which $\rho \sim 0.0$.
From this, we find that there is no valid binary lens solution with small $\rho$.
This is because although we find two solutions with better $\chi^2$ relative to the single lens model, the best fit lens-source relative proper motions are $\murel=177\, \rm mas\, yr^{-1}$ and $\murel=583\,\rm mas\, yr^{-1}$ for the $\rho=0.0086$ and $\rho=0.0018$ solutions, respectively.
These are very large (as anticipated in the previous paragraph) to the extent that they are unphysical.
One of the two solutions (for $\rho=0.0086$) is the binary system composed of a primary star $M_{\rm L,1}=1.69 \pm 9.17\ M_\odot$ and a planet $M_{\rm L,2}=1.21 \pm 6.54\ M_{\rm Jupiter}$ with their projected separation 9.3 AU, while for the other solution (for $\rho=0.0018$) it is the binary system composed of a primary star $M_{\rm L,1}=5.55 \pm 11.26\ M_\odot$ and a planet $M_{\rm L,2}=5.00 \pm 10.15\ M_{\rm Jupiter}$ with the separation 14.7 AU, and these binaries are respectively located at 3.7 kpc and 1.6 kpc.
The very large proper motion is due to large $\thetae$, while $\te$ is clearly determined from dense ground-based observations, as mentioned in the previous paragraph.
Moreover, the $\chi^2$ of the \textit{Spitzer} data for the two binary lens models becomes worse.

\section{CONCLUSION}

We analyzed the single lens event OGLE-2015-BLG-1482 simultaneously observed from two ground-based surveys and from \textit{Spitzer}.
The \textit{Spitzer} data exhibit the finite-source effect due to the passage of the lens directly over the surface of the source star as seen from \textit{Spitzer}.
Thanks to the finite-source effect and the simultaneous observation from Earth and \textit{Spitzer},  we were able to measure the mass of the lens.
From this analysis, we found that the lens of OGLE-2015-BLG-1482 is a very low-mass star with the mass $0.10 \pm 0.02 \ M_\odot$ or a brown dwarf with the mass $55\pm 9 \ M_J$, which are respectively located at $D_{\rm LS} = 0.80 \pm 0.19\ \textrm{kpc}$ and $ D_{\rm LS} = 0.54 \pm 0.08\ \textrm{kpc}$, and thus it is the first isolated low-mass object located in the Galactic bulge.
The degeneracy between the two solutions is very severe ($\delcs = 0.3$).

The fundamental reason for the degeneracy is that the finite-source effect is seen only in a single data point from \textit{Spitzer} and this data point has the finite-source effect function $B(z) = A_{\rm obs} u > 1$, where $z = u/\rho$.
We showed that whenever $B(z) > 1$, there are two solutions for $z$ and hence for $\rho = u/z$.
Because the $\rho$ degeneracy can be resolved only by relatively high cadence observations around the peak,  while the \textit{Spitzer} cadence is typically $\sim 1\ {\rm day}^{-1}$, we expect that events where the finite-source effect is seen only in the \textit{Spitzer} data may frequently exhibit the $\rho$ degeneracy.

In the case of OGLE-2015-BLG-1482, the lens-source relative proper motion for the low-mass star is $\murel = 9.0 \pm 1.9\ \textrm{mas\ yr$^{-1}$}$, while for the brown dwarf it is $5.5 \pm 0.5\ \textrm{mas\ yr$^{-1}$}$.
Hence, the severe degeneracy can be resolved within $\sim 10\ \rm yr$ from direct lens imaging by using next-generation instruments with high spatial resolution.

\acknowledgments
Work by S.-J. Chung was supported by the KASI (Korea Astronomy and Space Science Institute) grant 2017-1-830-03.
Work by W.Z. and A.G. was supported by JPL grant 1500811.
Work by C.H. was supported by the Creative Research Initiative Program (2009-0081561) of National Research Foundation of Korea.
This research has made use of the KMTNet system operated by KASI and the data were obtained at three host sites of CTIO in Chile, SAAO in South Africa, and SSO in Australia.
The OGLE has received funding from the National Science Centre, Poland, grant MAESTRO 2014/14/A/ST9/00121 to A.U. 
The OGLE Team thanks Professors M. Kubiak, G. Pietrzy{\`n}ski, and {\L}.Wyrzykowski for their contribution to the collection of the OGLE photometric data over the past years.

\appendix
\section{FINITE-SOURCE EFFECTS}

In the high-magnification limit, the magnification of a point source is $A_{\rm ps} \simeq 1/u$.
Considering the limb darkening effect in high-magnification events, the ratio between the magnifications with and without the finite-source effect is expressed as
\begin{equation}
B(z) = u{\int_0^\rho dr\,r \int_0^{2\pi}d\theta A_{\rm ps}(|{\bf u} + r \hat {\bf n}(\theta)|) 
\left[1 - \Gamma(1 - 1.5\sqrt{1-(r/\rho)^2})\right]\over \pi\rho^2},
\end{equation}
where $u$ is the normalized separation between the lens and the center of the source, $\hat {\bf n}(\theta) = (\cos\theta,\sin\theta)$, and $r$ and $\theta$ are the position vector and the position angle of a point on the source surface with the respect to the source center, respectively (see \citealt{gould94a, gould08}).
Changing variables to $x=r/\rho$,
\begin{eqnarray}
B(z) & = & z{\int_0^1 dx\,x \int_0^{2\pi}d\theta |{\bf z} + x \hat {\bf n}(\theta)|^{-1} 
\left[1 - \Gamma(1 - 1.5\sqrt{1-x^2})\right] \over \pi} \nonumber \\
      &  =  & {\int_0^1 dx\,x \left[1 - \Gamma(1 - 1.5\sqrt{1-x^2})\right] \int_0^{2\pi}d\theta \left[1 + 2(x/z)\cos\theta + (x/z)^2\right]^{-1/2}\over \pi}.
\end{eqnarray}
Here we change $x/z = Q$, and Taylor expand the first factor in the integrand,
\begin{eqnarray}
(1 + 2Q\cos\theta + Q^2)^{-1/2} & = &1 - {1\over 2}(2Q\cos\theta + Q^2) + {3\over 8}(2Q\cos\theta + Q^2)^2 -{15\over 48}(2Q\cos\theta + Q^2)^3 + {105\over 384}(2Q\cos\theta + Q^2)^4 + \dots \nonumber
\end{eqnarray}
Then, keeping terms only to $Q^4$,
\begin{gather*}
\int_0^{2\pi}d\theta (1 + 2Q\cos\theta + Q^2)^{-1/2}  = \\
\int_0^{2\pi}d\theta \left[1 -Q\cos\theta + \left(-{1\over2} + {3\over2}\cos^2\theta\right)Q^2 + {1\over2}\cos\theta\left(3-5\cos^2\theta\right)Q^3 + \left({3\over8} -{15\over 4}\cos^2\theta
+{35\over 8}\cos^4\theta\right)Q^4\right]
\end{gather*}
\begin{eqnarray*}
& = &2\pi\left[1 + \left({3\over 4}-{1\over 2}\right)Q^2 + \left({3\over 8} - {15\over 8} + {105\over 64}\right)Q^4\right] \\
& = & 2\pi\left(1 + {1\over 4}Q^2 + {9\over 64}Q^4\right) .
\end{eqnarray*}
With $y\equiv x^2$, Equation (A2) becomes
\begin{align*}
B(z) & = \int_0^1 dy \left[1 - \Gamma(1 - 1.5(1-y)^{1/2})\left(1 + {1\over 4}{y\over z^2} + {9\over 64}{y^2\over z^4}\right)\right]  \\
      & = \int_0^1 dy \left(1 + {1\over 4}{y\over z^2} + {9\over 64}{y^2\over z^4}\right)- \Gamma\int_0^1 dy\left[\left(1-1.5(1-y)^{1/2}\right)\left(1 + {1\over 4}{y\over z^2} + {9\over 64}{y^2\over z^4}\right)\right].
\end{align*}
Noting that $\int_0^1 x^a (1-x)^b = a! b!/(a+b+1)!$, we get
\begin{align*}
B(z) &= 1 + {1\over 8}{1\over z^2} + {3\over 64}{1\over z^4} -\Gamma\left[1 + {1\over 8}{1\over z^2} + {3\over 64}{1\over z^4} 
- 1.5\left({2\over 3} +  {1\over 15}{1\over z^2} + {3\over{4\times 35}}{1\over z^4}\right)\right] \\
      & = 1 + {1\over 8}{1\over z^2} + {3\over 64}{1\over z^4} -\Gamma\left({1\over 40}{1\over z^2} + {33\over {64\times35}}{1\over z^4}\right).
\end{align*}
Then, we finally get
\begin{equation}
B(z) = 1 + {1\over 8}\left(1 - {\Gamma\over 5}\right){1\over z^2} + {3\over 64}\left(1 - {11\over 35}\Gamma\right){1\over z^4}.
\end{equation}

%====================================================================
%========// Figures //===================================================
%====================================================================
%Figure 1 --------------------------------------------------------------
\begin{figure*}
\centering
\includegraphics[width=150mm]{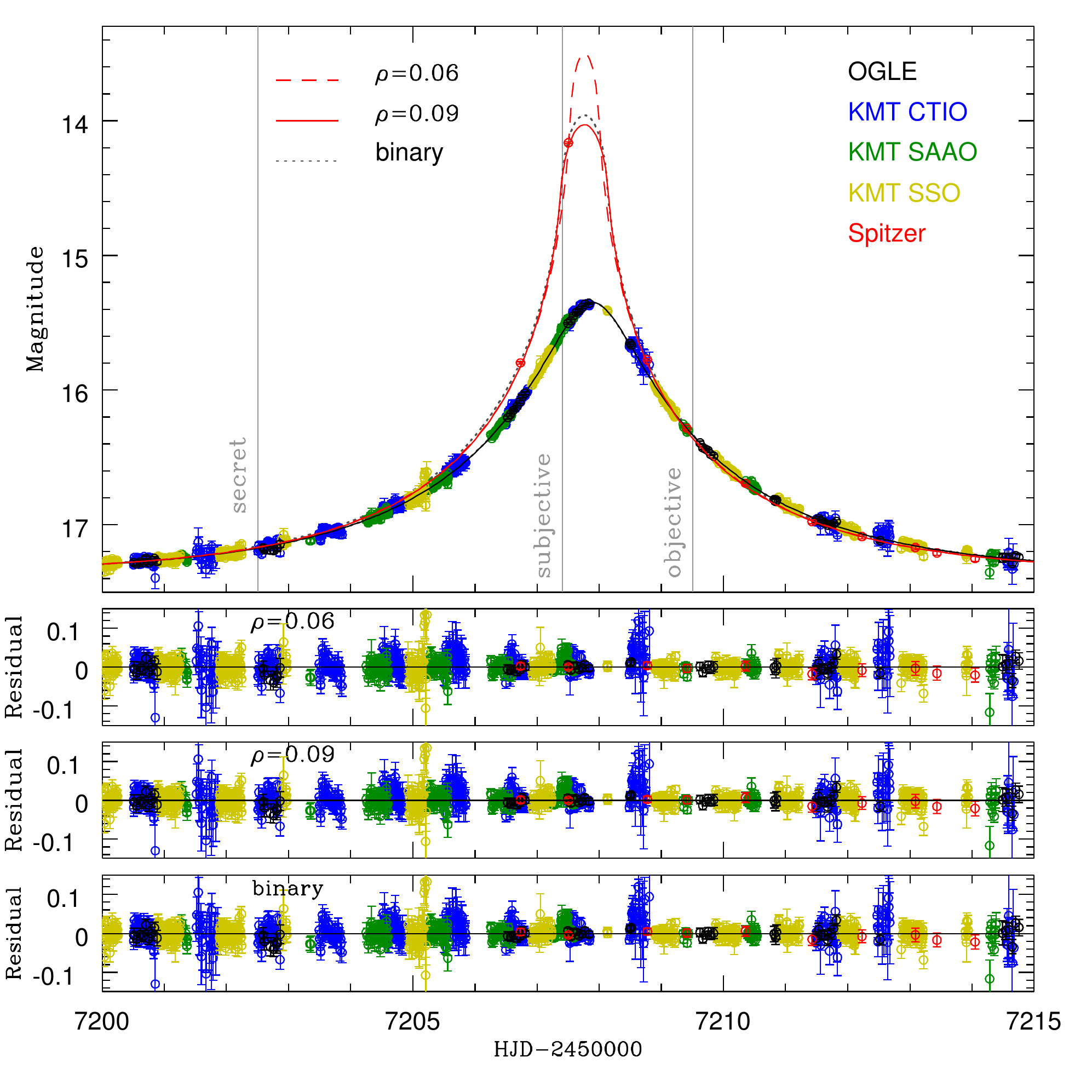}
\caption{\label{fig:one}
Light curves of the best-fit single lens model for OGLE-2015-BLG-1482.
The light curves of the best-fit binary lens model are also shown in the figure, and they are drawn by a dark grey dotted line.
The finite-source effect is constrained by only one single \textit{Spitzer} data point, which leads to two models with different $\rho$ values.
The grey vertical lines represent the times when secret, subjective, and objective alerts were issued.
}
\end{figure*}

%Figure 2 --------------------------------------------------------------
\begin{figure}
\centering
\includegraphics[width=150mm]{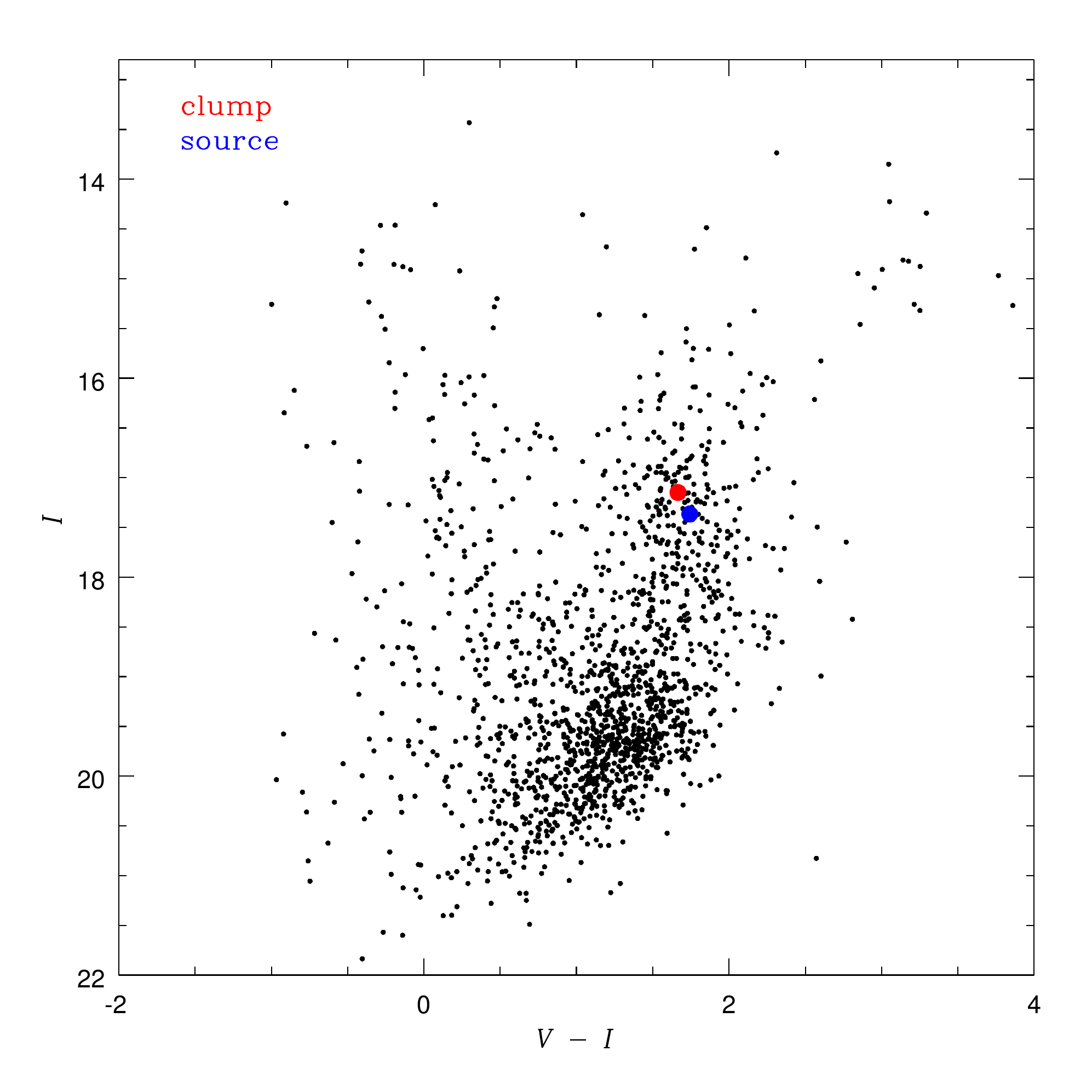}
\caption{\label{fig:two}
Color-magnitude diagram (CMD) of stars in the observed field.
The field stars are taken from KMTNet CTIO data.
We note that there exists an offset between the instrumental magnitudes of OGLE and KMTNet as $I_{\rm kmt} - I_{\rm ogle} = 0.045$ mag.
The red and blue circles mark the centroid of the red clump giant and the microlensed source star, respectively.
}
\end{figure}

%Figure 3 --------------------------------------------------------------
\begin{figure}
\centering
\includegraphics[width=150mm]{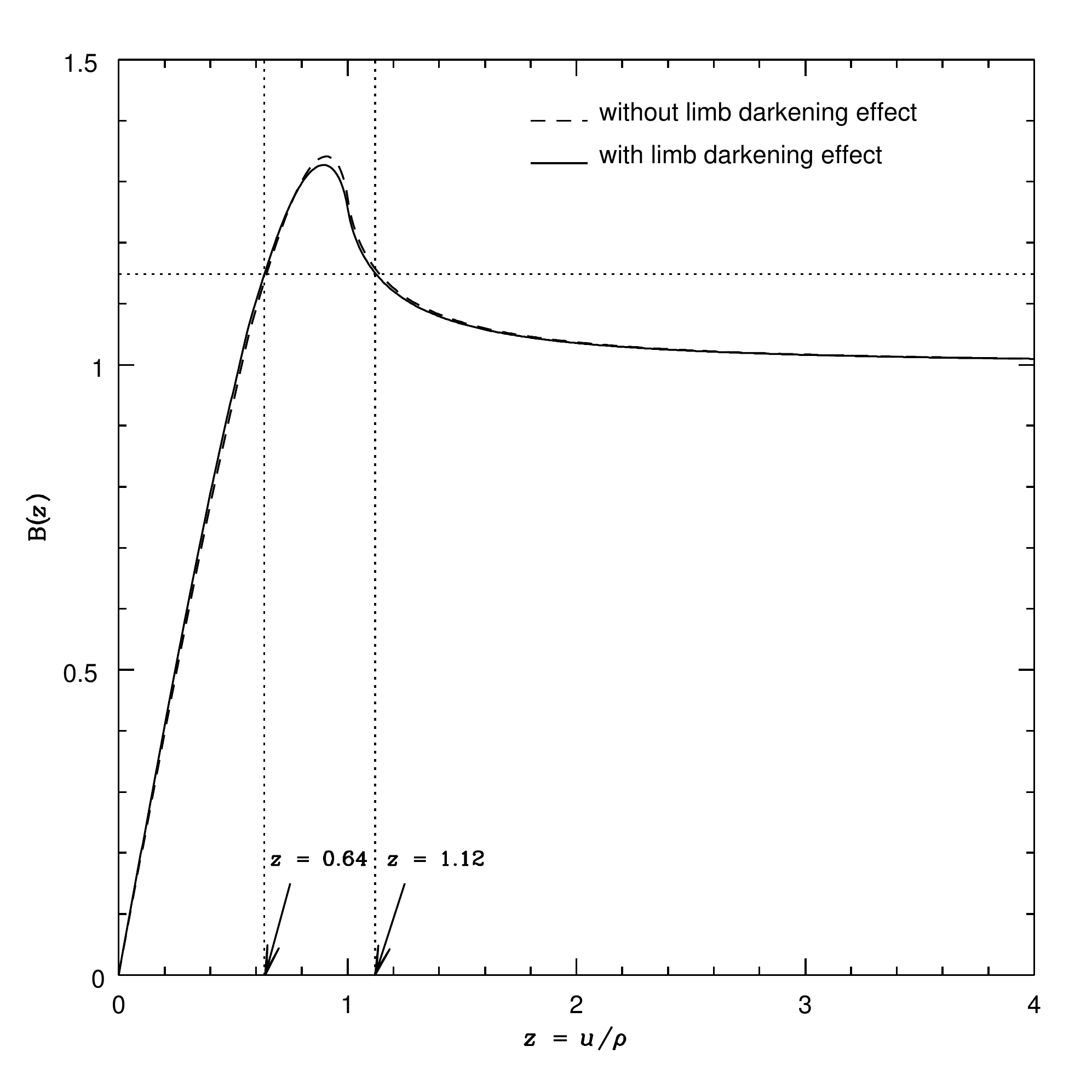}
\caption{\label{fig:three}
Ratio between the actual magnification including finite-source effects to the magnification of a point source, $B(z)$, as a function of $z\equiv u/\rho$, i.e., the ratio of the lens-source projected separation to the source radius.
In contrast to \citet{gould94a} from which this figure is adapted, we show the magnification both with (solid) and without (dashed) limb darkening.  
The horizontal dotted line indicates $B(z) = 1.15$.
}
\end{figure}

%Figure 4 --------------------------------------------------------------
\begin{figure}
\centering
\includegraphics[width=150mm]{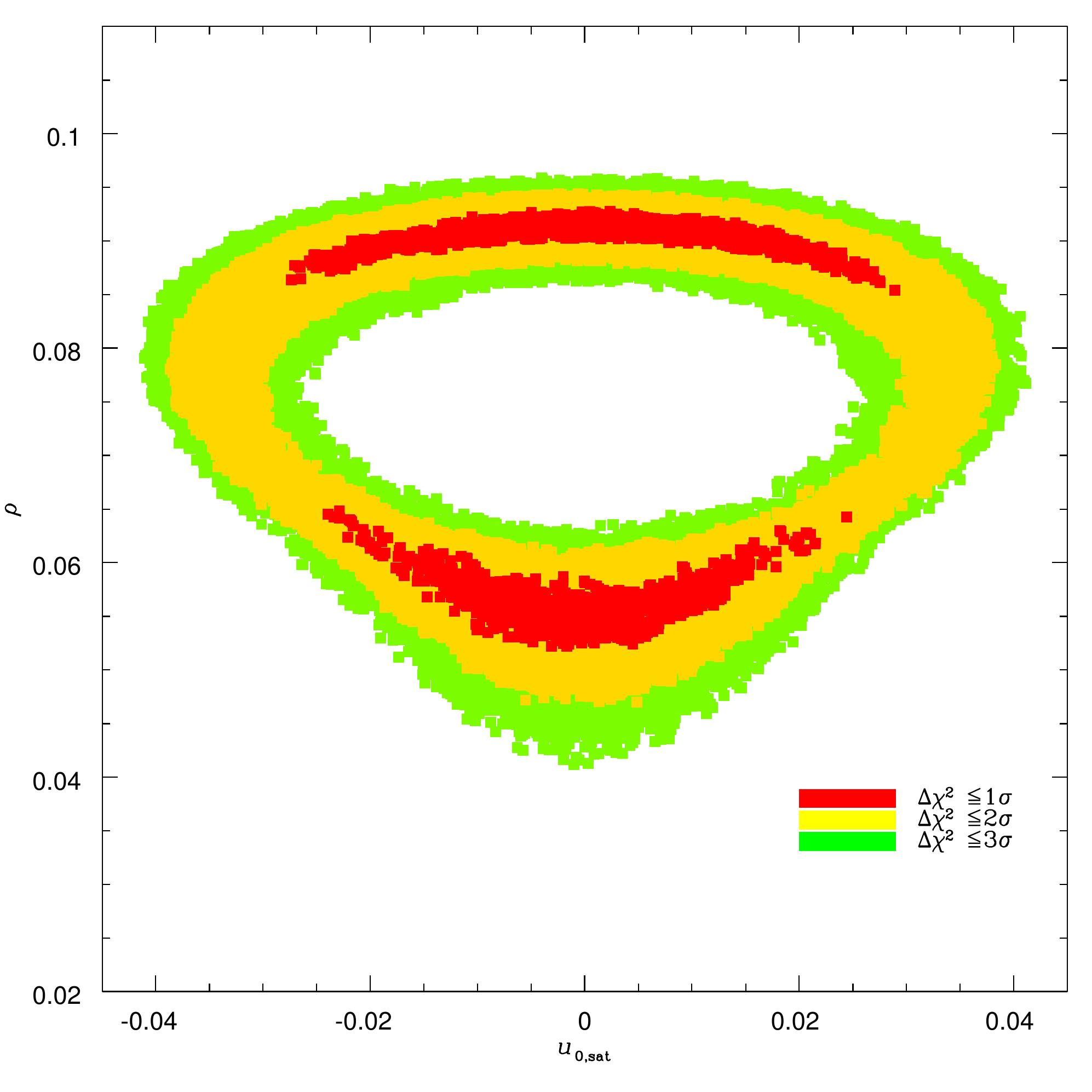}
\caption{\label{fig:four}
$\chi^2$ distribution of $\uos$ versus $\rho$ from the MCMC chains of four degenerate solutions in Table 1.
}
\end{figure}

%====================================================================
%========// Tables //===================================================
%====================================================================
\clearpage
\begin{turnpage}
\begin{deluxetable}{cccrcccccccccccc}
\tablewidth{0pt}
%\tabletypesize{\scriptsize}
\tablecaption{Best-fit parameters.\label{tbl-one}}
\tablehead{
&& \multicolumn{10}{c}{Fit parameters} \\
\cline{3-12}\\
Solutions   && $\chi^2/\rm{dof}$ & $t_{0} (\rm HJD')$ & $u_{0}$ & $\te$ (days) & $\rho(10^{-2})$ & $\pien$  & $\piee$ & $f_{s,ogle}$ & $f_{b,ogle}$
}
\startdata
 \multirow{2}{*}{$(+, 0)$}  &&  8360.63/8367    & $7207.893 \pm 0.001$   &  $0.160 \pm 0.002$    & $4.265 \pm 0.021$  &  $5.55 \pm 1.10$ & $-0.1288 \pm 0.0169$ & $0.0346 \pm  0.0016$ & $1.790 \pm 0.015$ & $-0.004 \pm 0.015$\\                                    
                                         &&  8360.92/8367    & $7207.893 \pm 0.001$   &  $0.165 \pm 0.002$    & $4.258 \pm 0.022$  &  $9.16\pm 0.60$ & $-0.1342\pm 0.0189$   & $0.0349 \pm 0.0017$  & $1.794 \pm 0.015$ & $-0.009 \pm 0.015$ \\\\     
\multirow{2}{*}{$(-, 0)$}   &&  8360.95/8367    & $7207.893 \pm 0.001$   &  $-0.160 \pm 0.002$    & $4.265 \pm 0.021$  &  $5.55 \pm 1.08$ & $0.1309 \pm 0.0163$ & $0.0159 \pm 0.0017$ & $1.790 \pm 0.015$ & $-0.005 \pm 0.015$ \\
                                         &&  8361.15/8367    & $7207.893 \pm 0.001$   &  $-0.164 \pm 0.002$    & $4.262 \pm 0.022$  &  $9.10\pm 0.59$ & $0.1342 \pm 0.0188$ & $0.0155 \pm 0.0018$ & $1.791 \pm 0.015$ & $-0.005 \pm 0.015$ \\                                                                                       
\enddata                                                    
\tablecomments{$(+,0)$ indicates $\uoe > 0$ and $\uos \simeq 0$. $\rm HJD'$ is ${\rm HJD} - 2450000$.
}
\end{deluxetable}
%\end{turnpage}

%\clearpage
%\begin{turnpage}
\begin{deluxetable}{cccccccccccccccc}
\tablewidth{0pt}
\tablecaption{Best-fit parameters for remodeling.\label{tbl-two}}
\tablehead{
 && \multicolumn{10}{c}{Fit parameters} \\
\cline{3-12}\\
Solutions   && $\chi^2/\rm{dof}$ & $t_{0} (\rm HJD')$ & $u_{0}$ & $\te$ (days) & $\rho(10^{-2})$ & $\pien$  & $\piee$ & $f_{s,ogle}$ & $f_{b,ogle}$
 }
\startdata
\multirow{2}{*}{$(+, 0)$}  &&  8361.00/8367    & $7207.893 \pm 0.002$   &  $0.160 \pm 0.002$    & $4.268 \pm 0.031$  &  $5.64 \pm 1.19$ & $-0.1303 \pm 0.0179$ & $0.0344 \pm 0.0021$  & $1.787 \pm 0.022$  & $-0.002 \pm 0.022$ \\
                                        &&  8361.20/8367    & $7207.893\pm 0.002$    &  $0.165 \pm 0.002$    & $4.259 \pm 0.032$  &  $9.10\pm 0.68$  & $-0.1356 \pm 0.0210$ & $0.0345 \pm 0.0021$ & $1.793 \pm 0.023$  & $-0.008 \pm 0.023$ \\\\            
\multirow{2}{*}{$(-, 0)$}  &&  8361.08/8367    & $7207.893 \pm 0.002$   &  $-0.160 \pm 0.002$    & $4.264 \pm 0.032$  &  $5.77 \pm 1.13$ & $0.1254 \pm 0.0167$ & $0.0162 \pm 0.0023$ & $1.790 \pm 0.022$  & $-0.005 \pm 0.022$ \\
                                       &&  8361.27/8367    & $7207.893 \pm 0.002$   &  $-0.164 \pm 0.002$    & $4.262 \pm 0.031$  &  $9.10\pm 0.64$  & $0.1382 \pm 0.0200$ & $0.0153 \pm 0.0023$ & $1.790 \pm 0.022$  & $-0.005 \pm 0.022$ \\      
\enddata
\tablecomments{This is the result of remodeling included $(f_s, f_b)_{spitzer}$ as chain parameters                             
}
\end{deluxetable}
\end{turnpage}
%\clearpage

\begin{table}
\center
\caption{Ranges of recognizable finite-source effects for a single data point.\label{tbl-three}}
\begin{threeparttable}
\begin{tabular}{lccc}
\cline{1-4}
\\
Range (B)  &  $0<B<0.95$ &  $1.05<B<1.34$  &  $1.34 > B^{a} > 1.05$ \\ 
Range (z)  &  $0 < z < 0.51$ &  $0.57 < z < 0.91$  & $ 0.91 < z < 1.70$ \\ 
Length (z) &  0.51 & 0.34 & 0.79 \\
$\rho$ solution    & single & two (higher $\rho$)  & two (lower $\rho$) \\
\\
\cline{1-4}
\end{tabular}    
\begin{tablenotes}
\item[a] The range $1.34 > B > 1.05$ represents the decreasing range of $B(z)$ curve (i.e., from $B=1.34$ (peak) to $B=1.05$), as shown in Figure 3.
\end{tablenotes}
\end{threeparttable}                   
\end{table}
%=====================================================================


\begin{thebibliography}{}
\frenchspacing

\bibitem[Alard \& Lupton(1998)]{alard98}
Alard, C., \& Lupton, R.\ H.\ 1998, \apj, 503, 325

\bibitem[Albrow et al.(2001)]{albrow01}
Albrow, M.\ D., An, J., Beaulieu, J.-P., et al.\ 2001, \apj, 556, L113

\bibitem[Albrow et al.(2009)]{albrow09}
Albrow, M.\ D., Horne, K., Bramich, D.\ M., et al.\ 2009, \mnras, 397, 2099

\bibitem[An et al.(2002)]{an02a}
An, J.\ H., Albrow, M.\ D., Beaulieu J.-P., et al.\ 2002, \apj, 572, 521

\bibitem[Bachelet et al. (2012)]{bachelet12}
Bachelet, E., Fouqu{\'e}, P., Han, C., et al.\ 2012, \aap, 547, 55

\bibitem[Bensby et al. (2011)]{bensby11}
Bensby, T., Ad\'{e}n, D., Mel\'{e}ndez, et al.\ 2011 \aap, 533, 134

\bibitem[Bessell \& Brett(1988)]{bessell88}
Bessell, M.\ S., \& Brett, J.\ M.\ 1988, \pasp, 100, 1134

\bibitem[Bozza et al.(2012)]{bozza12}
Bozza, V., Dominik, M., Rattenbury, N.\ J., et al.\ 2012, \mnras, 424, 902

\bibitem[Calchi Novati et al.(2015)]{calchinovati15}
Calchi Novati, S., Gould, A., Yee, J.\ C., et al.\ 2015, \apj, 814, 92

\bibitem[Choi et al.(2013)]{choi13}
Choi, J.-Y., Han, C., Udalski, A., et al.\ 2013, \apj, 768, 129

\bibitem[Claret(2000)]{claret00}
Claret, A.\ 2000, \aap, 363,1081

\bibitem[Claret \& Bloemen(2011)]{claret11}
Claret, A.,\& Bloemen, S.\ 2011, \aap, 529, A75

\bibitem[Close et al.(2013)]{close13}
Close, L.\ M., Males, J.\ R., Morzinski, K., et al.\ 2013, \apj, 774, 94

\bibitem[Deleuil et al.(2008)]{deleuil08}
Deleuil, M., Deeg, H.\ J., Alonso, R., et al.\ 2008, \aap, 491, 889

\bibitem[D\'{i}az et al.(2013)]{diaz13}
D\'{i}az, R.\ F., Damiani, C., Deleuil, M., et al.\ 2013, \aap, 551, L9

\bibitem[Dong et al.(2007)]{dong07}
Dong, S., Udalski, A., Gould, A., et al.\ 2007, \apj, 664, 862

\bibitem[Gaudi et al. (2002)]{gaudi02}
Gaudi, B.\ S., Albrow, M.\ D., An, J., et al.\ 2002 \apj, 566, 463

\bibitem[Gould (1992)]{gould92}
Gould, A.\ 1992 \apj, 392, 442

\bibitem[Gould \& Loeb(1992)]{gould92b}
Gould, A., \& Loeb A.\ 1992 \apj, 396, 104

\bibitem[Gould (1994a)]{gould94a}
Gould, A.\ 1994a \apj, 421, L71

\bibitem[Gould (1994b)]{gould94b}
Gould, A.\ 1994b \apj, 421, L75

\bibitem[Gould (1995)]{gould95}
Gould, A.\ 1995 \apj, 441, L21

\bibitem[Gould (1997)]{gould97}
Gould, A.\ 1997 \apj, 480, 188

\bibitem[Gould (2000)]{gould00}
Gould, A.\ 2000 \apj, 542, 785

\bibitem[Gould (2004)]{gould04}
Gould, A.\ 2004 \apj, 606, 319

\bibitem[Gould (2008)]{gould08}
Gould, A.\ 2008 \apj, 681, 1593

\bibitem[Gould et al.(2009)]{gould09}
Gould, A., Udalski, A., Monard, B., et al.\ 2009 \apj, 698, L147

\bibitem[Gould \& Yee(2012)]{gould12}
Gould, A. \& Yee, J.\ C.\ 2012 \apj, 755, L17

\bibitem[Gould \& Yee(2013)]{gould13}
Gould, A. \& Yee, J.\ C.\ 2013 \apj, 764, 107

\bibitem[Gould et al.(2014)]{gould14}
Gould, A., Carey, S., \& Yee, J.\ 2014, Spitzer Proposal ID\#11006

\bibitem[Griest \& Safizadeh(1998)]{griest98}
Griest, K. \& Safizadeh, N.\ 1998 \apj, 500, 37

\bibitem[Han \& Chang(2003)]{han03}
Han, C. \& Chang, H.-Y.\ 2003, \mnras, 338, 637

\bibitem[Han et al.(2016)]{han16}
Han, C., Jung, Y.\ K., Udalski, A., et al., 2016, \apj, 822, 75

\bibitem[Henderson et al.(2014)]{henderson14}
Henderson, C.\ B., Gaudi, B.\ S., Han, C. et al.\ 2014 , \apj, 794, 52

\bibitem[Johnson et al.(2011)]{johnson11}
Johnson, J.\ A., Apps, K., Gazak, J.\ Z., et al.\ 2011 \apj, 730, 79

\bibitem[Jung et al.(2015)]{jung15}
Jung, Y.\ K., Udalski, A., Sumi, T., et al.\ 2015 \apj, 798, 123

\bibitem[Kervella et al.(2004)]{kervella04}
Kervella, P., Th\'{e}venin, F., Di Folco, E., S\'{e}gransan, D.\ 2004, \aap, 426, 297

\bibitem[Kervella \& Fouqu\'{e}(2008)]{kervella08}
Kervella, P., \& Fouqu\'{e}, P.\ 2008, \aap, 491, 855

\bibitem[Kim et al.(2016)]{kim16}
Kim, S.-L., Lee, C.-U., Park, B.-G., et al.  2016, JKAS, 49, 37

\bibitem[Lafreni\`{e}re et al.(2007)]{lafreniere07}
Lafreni\`{e}re, D., Doyon, R., Marois, C., et al.\ 2007, \apj, 670, 1367

\bibitem[Laureijs et al.(2011)]{laureijs11}
Laureijs, R., Amiaux, J., Arduini, S.,  et al.\ 2011, arXiv:1110.3193

\bibitem[Luhman (2012)]{luhman12}
Luhman, K.\ L.\ 2012, \araa, 50, 65

\bibitem[McGregor et al.(2012)]{mcgregor12}
McGregor, P.\ J., Bloxham, G.\ J., Boz, R., et al.\ 2012, \procspie, 8446, 84461I

\bibitem[Moutou et al.(2013)]{moutou13}
Moutou, C., Bonomo, A.\ S., Bruno, G., et al.\ 2013, \aap, 558, L6

\bibitem[Nataf et al.(2013)]{nataf13}
Nataf, D.\ H., Gould, A., Fouqu\'{e'}, P. et al.\ 2013, \apj, 769, 88

\bibitem[Park et al.(2013)]{park13}
Park, H., Udalski, A., Han, C., et al.\ 2013, \apj, 778, 134

\bibitem[Park et al.(2015)]{park15}
Park, H., Udalski, A., Han, C., et al.\ 2015, \apj, 805, 117

\bibitem[Refsdal (1966)]{refsdal66}
Refsdal, S.\ 1966 \mnras, 134, 315

\bibitem[Sahlmann et al.(2011)]{sahlmann11}
Sahlmann, J., S\'{e}gransan, D., Queloz, D., et al.\ 2011, \aap, 525, A95

\bibitem[Shin et al.(2012)]{shin12a}
Shin, I.-G., Choi, J.-Y., Park, S.-Y., et al.\ 2012a, \apj, 746, 127

\bibitem[Shin et al.(2012)]{shin12b}
Shin, I.-G., Han, C., Gould, A., et al.\ 2012b, \apj, 760, 116

\bibitem[Shin et al.(2016)]{shin16}
Shin, I.-G., Ryu, Y.-H., Udalski, A., et al.\ 2016, JKAS, 49, 73

\bibitem[Siverd et al.(2012)]{siverd12}
Siverd, R.\ J., Beatty, T.\ G., Pepper, J., et al.\ 2012, \apj, 761, 123

\bibitem[Skowron et al.(2015)]{skowron15}
Skowron J., Shin, I.-G., Udalski, A., et al.\ 2015, \apj, 804, 33

\bibitem[Spergel et al.(2015)]{spergel15}
Spergel, D., Gehrels, N., Baltay, C., et al.\ 2015, arXiv:1503.03757v2

\bibitem[Street et al.(2013)]{street13}
Street, R.\ A., Choi, J.-Y., Tsapras, Y., et al.\ 2013, \apj, 763, 67

\bibitem[Sumi et al.(2011)]{sumi11}
Sumi, T., Kamiya, K., Bennett, D.\ P., et al.\ 2011, Nature, 473, 349

\bibitem[Udalski(2003)]{udalski03}
Udalski, A.\ 2003, AcA, 53, 291

\bibitem[Udalski et al.(2015)]{udalski15}
Udalski, A., Yee, J.\ C., Gould, A., et al.\ 2015, \apj, 799, 237

\bibitem[Yee et al.(2009)]{yee09}
Yee, J.\ C., Udalski, A., Sumi, T., et al.\ 2009, \apj, 703, 2082

\bibitem[Yee et al.(2012)]{yee12}
Yee, J.\ C., Shvartzvald, Y., Gal-Yam, A., et al.\ 2012, \apj, 755, 102

\bibitem[Yee (2013)]{yee13}
Yee, J.\ C.\ 2013, \apj, 770, L31

\bibitem[Yee et al.(2015a)]{yee15a}
Yee, J.\ C., Udalski, A., Calchi Novati, S.,  et al.\ 2015a, \apj, 802, 76

\bibitem[Yee et al.(2015b)]{yee15b}
Yee, J.\ C., Gould, A., Beichman, C.,  et al.\ 2015b, \apj, 810, 155

\bibitem[Yoo et al.(2004)]{yoo04}
Yoo, J., DePoy, D.\ L., Gal-Yam, A., et al.\ 2004, \apj, 603, 139

\bibitem[Zhu \& Gould(2016)]{zhu16b}
Zhu, W., \& Gould, A.\ 2016, JKAS, 49, 93

\bibitem[Zhu et al.(2016)]{zhu16}
Zhu, W., Calchi Novati, S., Gould, A., et al.\ 2016, \apj, 825, 60


\end{thebibliography}
\end{document}